\documentclass{moriond}

\IfFileExists{srcltx.sty}{\usepackage[active]{srcltx}}%

\def\be{\begin{equation}}
\def\ee{\end{equation}}
\def\bea{\begin{eqnarray}}
\def\eea{\end{eqnarray}}

\newcommand{\Photo}{\includegraphics[height=35mm]{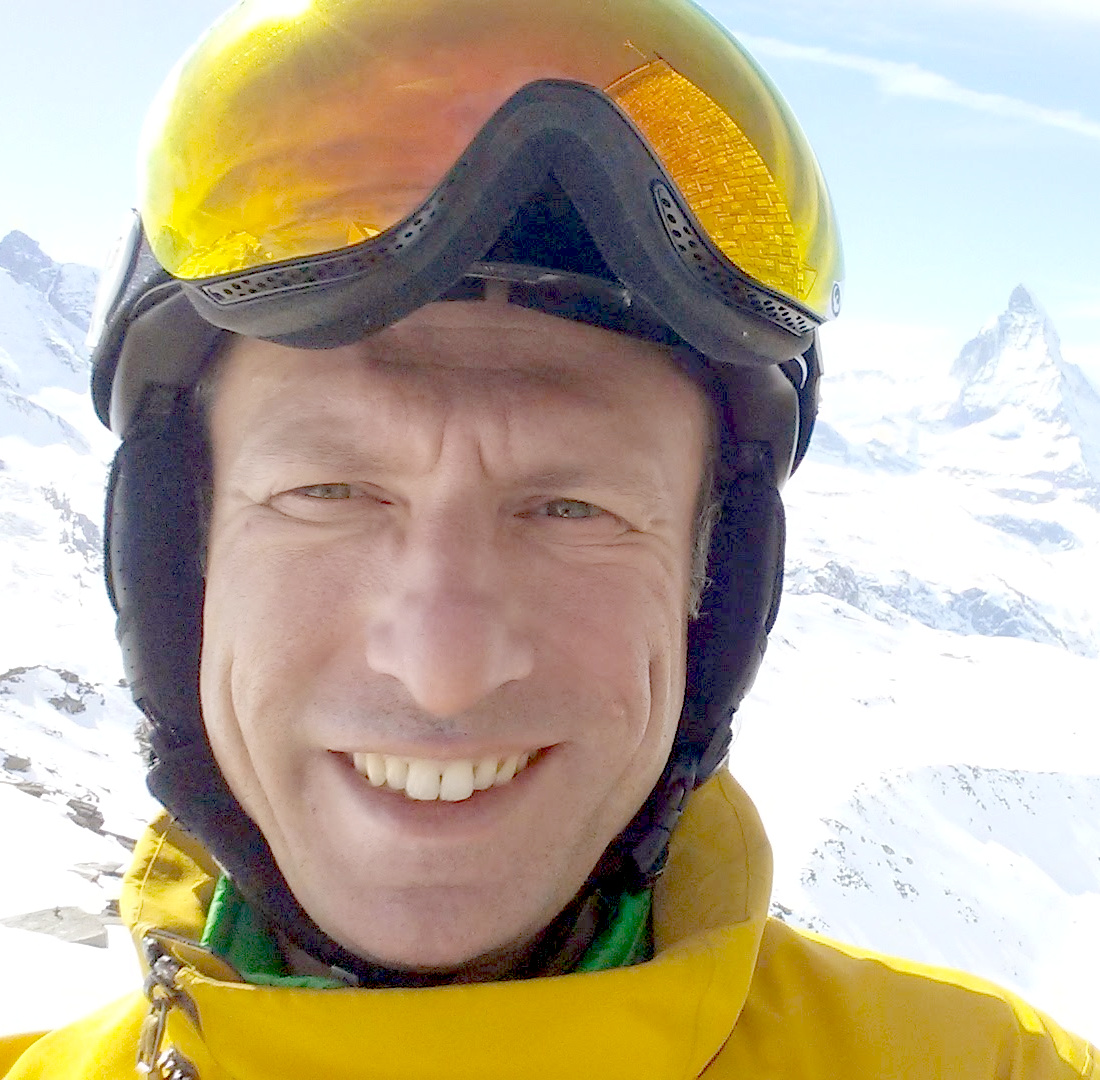}}

\begin{document}
\vspace*{4cm}
\title{Higgs relaxation and the matter-antimatter asymmetry of the universe}

\author{ Alexander Kusenko }

\address{Department of Physics and Astronomy, University of California, Los
Angeles, CA 90095-1547, USA\\Kavli Institute for the Physics and Mathematics of the Universe (WPI), University of Tokyo, Kashiwa, Chiba 277-8568, Japan}

\maketitle\abstracts{
The recent measurement of the Higgs boson mass implies a relatively slow rise of the Standard Model Higgs potential at large scales, and a possible second minimum at even larger scales. Consequently, the Higgs field may develop a large vacuum expectation value during inflation. The relaxation of the Higgs field from its large postinflationary value to the minimum of the effective potential represents an important stage in the evolution of the universe. During this epoch, the time-dependent Higgs condensate can create an effective chemical potential for the lepton number, leading to a generation of the lepton asymmetry in the presence of some large right-handed Majorana neutrino masses. Electroweak sphalerons redistribute this asymmetry between leptons and baryons. Higgs relaxation leptogenesis can explain the observed matter-antimatter asymmetry of the universe even if the Standard Model is valid up to the scale of inflation, and any new physics is suppressed by that high scale.
}

\section{Introduction}

During cosmological inflation, scalar fields, including the Higgs field can deviate from the minimum of the effective potential developing a large vacuum  expectation value (VEV).\cite{Bunch:1978yq}  The effect is most pronounced for those fields with flat directions  in the effective potentials, or with relatively shallow minima. The recent discovery of the Higgs mass of 125~GeV has allowed extrapolations of the Higgs potential to large scales, leading to the conclusion that, at large VEV, the potential becomes less steep, and that it may, in fact develop an instability.\cite{Espinosa:2007qp}  While this instability is probably  cured by some new physics at a high scale, there relatively slow rise of the Higgs potential implies that it was likely to have a large VEV at the end of inflation. 

Once the inflation is over, the Higgs mass must return to the minimum of the effective potential, and the epoch of Higgs relaxation can have observable consequences, such as the baryon asymmetry of the universe.\cite{Kusenko:2014lra,Pearce:2015nga}   Not only the Higgs relaxation, but also an axion relaxation\cite{Kusenko:2014uta} or a Majoron relaxation\cite{Ibe:2015nfa} can lead to the generation of the matter-antimatter asymmetry of the universe.  

\section{Higgs relaxation after inflation}
During de Sitter expansion, the Higgs field may be trapped in a quasistable second minimum or, alternatively, may develop a stochastic distribution of vacuum expectation values.  The two possibilities, arising from different assumptions regarding the new physics at a high scale, are shown in Fig.~1. The initial condition in the case of the false vacuum (IC-1) produces an equal initial value of the field on superhorizon scales.  At the end of inflation, reheating leads to a change in the effective potential, which eliminates the barrier, and the field starts rolling down the potential.  The other possible initial conditions (IC-2) leads to some stochastic distribution of values for the Higgs field.  In this case, some couplings between the Higgs field and the inflaton need to be introduced\cite{Kusenko:2014lra} to equalize the initial field values across superhorizon scales in order to avoid unacceptably large isocurvature perturbations.\cite{isocurvature}

\begin{figure}[ht!]
\centerline{\includegraphics[width=0.7\linewidth]{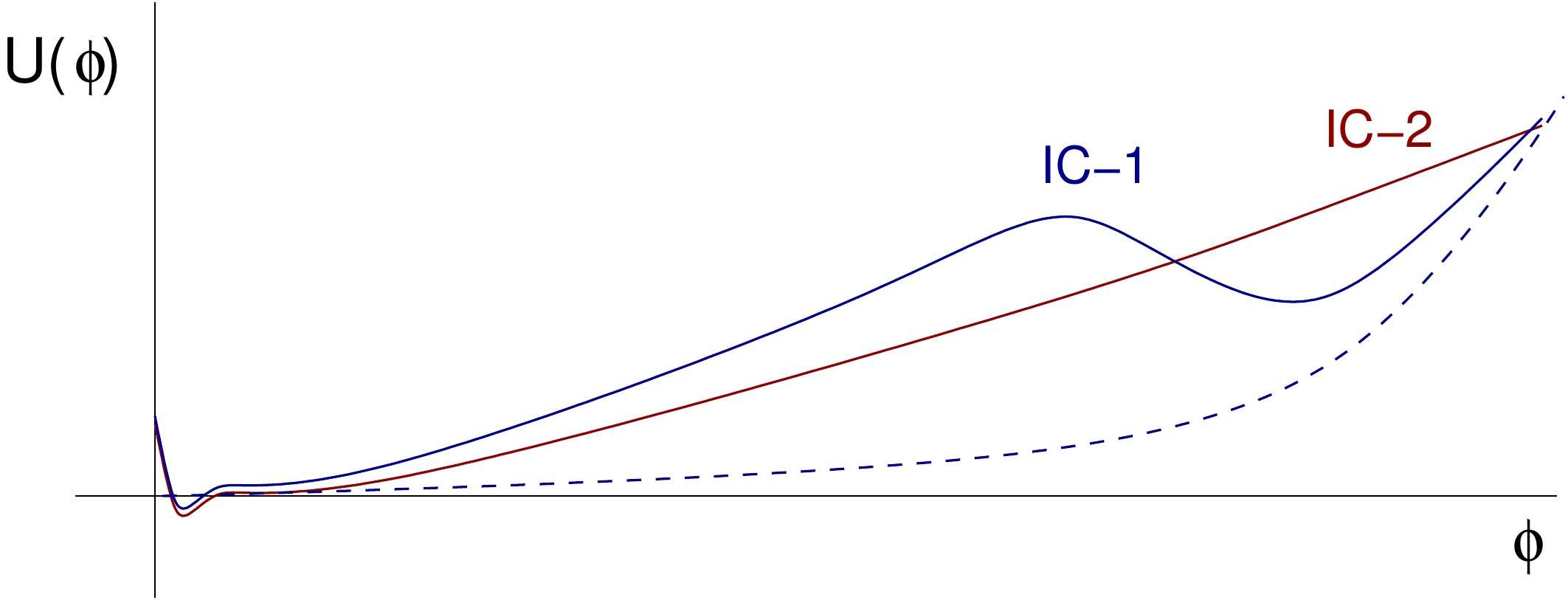}}
\caption[]{Two possibilities for the initial conditions, as discussed in the text.}
\label{fig:IC}
\end{figure}

\section{Leptogenesis via Higgs relaxation at high temperature}

The motion of the Higgs field can generate an effective chemical potential in plasma, if new physics at a scale $M_n$ gives rise to an operator 
\begin{equation}
{\cal O}_{6}  =  -\frac{1}{M_{n}^{2}}(\partial_{\mu}|\phi|^{2})\, j^{\mu},
\end{equation}
where 
$j^{\mu}$ is the fermion current whose zeroth component is the density of $(B+L)$.  This operator has been discussed in connection with spontaneous baryogenesis.\cite{Cohen:1987vi}  It breaks CPT spontaneously, and it generates an effective chemical potential in plasma leading to unequal energy levels between particles and antiparticles.

\begin{figure}[ht!]
\centerline{\includegraphics[width=0.7\linewidth]{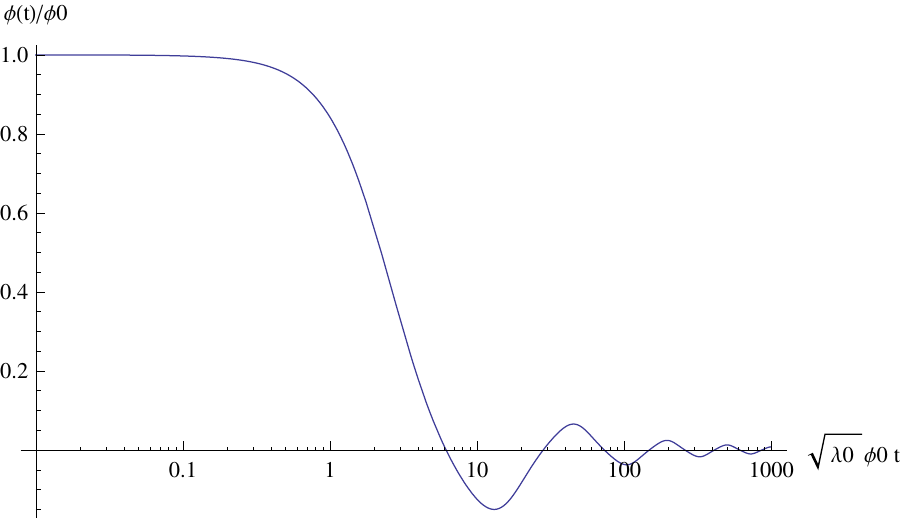}}
\caption[]{Time dependence of the Higgs VEV at the end of inflation.}
\label{fig:t}
\end{figure}

The amplitude of the oscillations of the Higgs field decreases with time (Fig.2), so that the first large swing dominates various effects of Higgs relaxation on plasma, and, during that time, the derivative $\partial_t \phi^\dag \phi$ is negative at all points in space.  Therefore, the shift in the energy levels of particles and antiparticles has the same sign everywhere.  

While the energy levels of particles and antiparticles are biased by the effective chemical potential, any process allowing the violation of baryon or lepton number leads to a particle-antiparticle asymmetry.  A heavy right-handed neutrino implied by the seesaw mechanism\cite{seesaw} can mediate such processes at high temperature via the diagrams shown in Fig.3.

\begin{figure}[ht!]
\centerline{\includegraphics[width=0.7\linewidth]{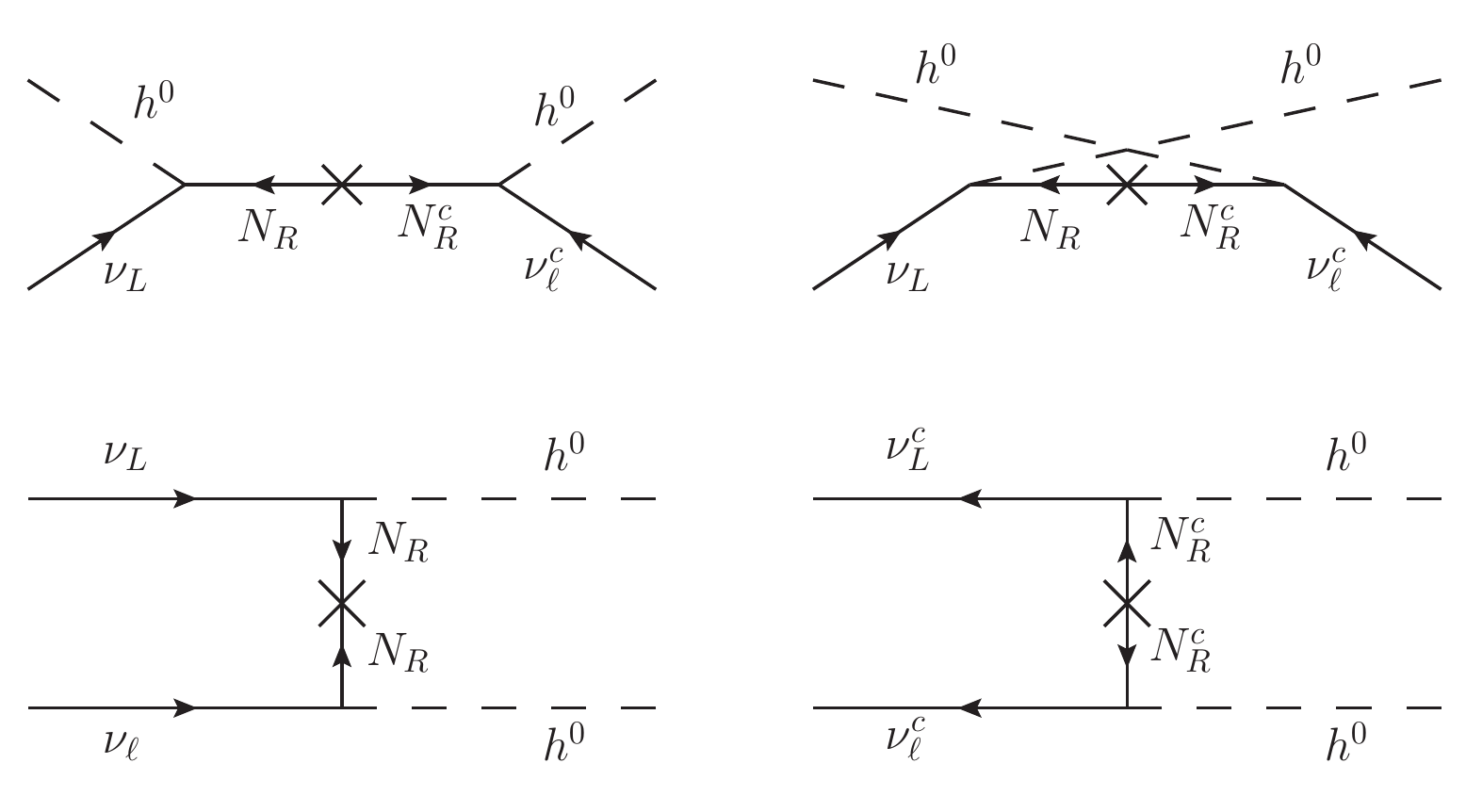}}
\caption[]{Lepton number (and $B-L$) violating diagrams with the heavy virtual neutrino exchange.}
\label{fig:IC1}
\end{figure}

A lepton asymmetry produced during the first large swing of the field undergoes partial washout in subsequent oscillations, and also after the oscillations subside.  The lepton asymmetry is redistributed between leptons and baryons by the sphaleron transitions, as in thermal leptogenesis.\cite{Fukugita:1986hr} The final baryon asymmetry is consistent with the observed matter-antimatter asymmetry of the universe for some reasonable values of 
parameters.\cite{Fukugita:1986hr}

\section{Leptogenesis via Higgs relaxation and particle production}

The scenario described above works for a relatively high reheat temperature, and the asymmetry is generated in plasma due to the presence of the effective chemical potential.  In a different regime of parameters, the more important effect is the (non-thermal) particle production by the time-dependent, oscillating Higgs field.\cite{Pearce:2015nga}  The asymmetry between particles and antiparticles can still be produced by the ${\cal O}_{6}$ operator, which shifts the energy levels of particles as compared to antiparticles.  The production rate is obtained by considering the Bogoliubov transformations for Majorana fermions in the presence of a time-dependent ${\cal O}_{6}$ operator, and a time-dependent effective mass\cite{Pearce:2015nga}.

This form of leptogenesis works particularly well when the Higgs condensate decays rapidly and at low reheat temperature.\cite{Pearce:2015nga}

\section{Conclusion}
Higgs relaxation at the end of inflation is a time when the matter-antimatter asymmetry could develop by way of non-thermal 
leptogenesis.\cite{Kusenko:2014lra,Pearce:2015nga}
Relaxation of an axion\cite{Kusenko:2014uta} or a Majoron\cite{Ibe:2015nfa} 
field offers an equally good opportunity for leptogenesis. This class of scenarios is different from other models 
of leptogenesis. In particular, the asymmetry can be
generated for reheat temperatures well below the righthanded
neutrino masses. This allows, in particular, for a supersymmetric
generalization of the model, in which the
problem of gravitino overproduction may not arise. Furthermore,
the final asymmetry does not depend on the parameters
of the neutrino mass matrix as in thermal leptogenesis,
and, therefore, a successful leptogenesis is possible even for
the neutrino masses above 0.2 eV, in which case thermal
leptogenesis is not possible due to an excessive washout.

\section{Acknowledgements} 
I thank L.~Pearce and L.~Yang for a fruitful and enjoyable collaboration.   
This work was supported by the U.S. Department of Energy Grant DE-SC0009937
and by the World Premier International Research Center Initiative
(WPI Initiative), MEXT, Japan.

\end{document}